\documentclass[runningheads]{llncs}

\usepackage{graphicx}
\usepackage{amssymb}
\usepackage{pgfplots}
\usepackage{tikz}
\usepackage{mathrsfs}
\usepackage{amsfonts}
\usepackage[ruled]{algorithm2e}
\usepackage{floatrow}
\usepackage{multirow}
\usepackage{bm}
\newfloatcommand{capbtabbox}{table}[][\FBwidth]
\SetCommentSty{mycommfont}
\usepackage[misc]{ifsym}

\newcommand{\etal}{{et al. }}
\newcommand{\ie}{i.e.}

\usepackage{hyperref}

\makeatletter
\newcommand{\printfnsymbol}[1]{%
  \textsuperscript{\@fnsymbol{#1}}%
}
\makeatother

\begin{document}
	%
	\title{Unsupervised Multi-Modality Registration Network based on Spatially Encoded Gradient Information}
	\titlerunning{Unsupervised MMRegNet based on Spatially Encoded Gradient Information}
	
    
    	\author{
    	Wangbin Ding \inst{1} \and
		Lei Li\inst{2,3} \and
		Liqin Huang\inst{1} \printfnsymbol{1} \and
		 Xiahai Zhuang\inst{2} \thanks{X Zhuang and L Huang are co-senior and corresponding authors: zxh@fudan.edu.cn; hlq@fzu.edu.cn. This work was funded by the National Natural Science Foundation of China (Grant No. 61971142), and Shanghai Municipal Science and Technology Major Project (Grant No. 2017SHZDZX01).}   } 
		\authorrunning{Ding et al.}
	\institute{College of Physics and Information Engineering, Fuzhou University, Fuzhou, China 
		\and School of Data Science, Fudan University, Shanghai, China
		\and School of Biomedical Engineering, Shanghai Jiao Tong University, Shanghai, China 
	}

\maketitle              

\setcounter{footnote}{0} 
\begin{abstract}
Multi-modality medical images can provide relevant or complementary information for a target (organ, tumor or tissue). 
Registering multi-modality images to a common space can fuse these comprehensive information, and bring convenience for clinical application. 
Recently, neural networks have been widely investigated to boost registration methods. 
However, it is still challenging to develop a multi-modality registration network due to the lack of robust criteria for network training. 
In this work, we propose a multi-modality registration network (MMRegNet), which can perform registration between multi-modality images. 
Meanwhile, we present spatially encoded gradient information to train MMRegNet in an unsupervised manner. 
The proposed network was evaluated on the public dataset from MM-WHS 2017. 
Results show that MMRegNet can achieve promising performance for left ventricle registration tasks. 
Meanwhile, to demonstrate the versatility of MMRegNet, we further evaluate the method using a liver dataset from CHAOS 2019. Our source code is publicly available\footnote{https://github.com/NanYoMy/mmregnet}.
		
\keywords{ Multi-Modality Registration \and Left Ventricle Registration \and Unsupervised Registration Network \and Gradient Information }

\end{abstract}
	
	\section{Introduction}
Registration is a critical technology to establish correspondences between medical images \cite{jour/mia/heinrich2012mind}. 
The study of registration algorithm enables tumor monitoring \cite{jour/bone/seeley2014co}, image-guided intervention \cite{jour/bbe/alam2018medical}, and treatment planning \cite{jour/eo/giesel2009image}. 
Multi-modality images, such as CT, MR and US, capture different anatomical information. 
Alignment of multi-modality images can help a clinician to improve the disease diagnosis and treatment \cite{jour/pm/fu2020deep}. For instance, Zhuang \cite{jour/pami/zhuang2018multivariate} registered multi-modality myocardium MR images to fuse complementary information for myocardial segmentation and scar quantification. 
Heinrich \etal \cite{conf/miccai/heinrich2013towards} performed registration of intra-operative US to pre-operative MR, which could aid image-guided neurosurgery. 

Over the last decades, various methods have been proposed to perform multi-modality image registration. 
The most common methods are based on statistical similarity metric, such as mutual information (MI) \cite{jour/tmi/maes1997multimodality}, normalized MI (NMI) \cite{jour/pr/studholme1999overlap} and spatial-encoded MI (SEMI) \cite{jour/tmi/zhuang2011nonrigid}.
Registrations are performed by maximizing these similarity metrics between the moved and fixed images. 
However, these metrics usually suffer from the loss of spatial information \cite{conf/miccai/qin2019unsupervised}.
Other common methods are based on invariant representation. 
Wachinger \etal \cite{jour/mia/wachinger2012entropy} presented the entropy and Laplacian image which are invariant structural representations across the multi-modality image, and registrations were achieved by minimizing the difference between the invariant representations.  
Zhuang \etal \cite{conf/iccis/zhuang2005medical} proposed the normal vector information of intensity image for registration, which obtained comparable performance to the MI and NMI. 
Furthermore, Heinrich \etal \cite{jour/mia/heinrich2012mind} designed a handcraft modality independent neighborhood descriptor to extract structure information for registrations.  
Nevertheless, these conventional methods solved the registration problem by iterative optimizing, which is not applicable for time-sensitive scenarios.

Recently, several registration networks, which could efficiently achieved registrations in an one-step fashion, have been widely investigated. 
Hu \etal \cite{jour/mia/hu2018weakly} proposed a weakly supervised registration neural network for multi-modality images by utilizing anatomical labels as the criteria for network training. 
Similarly, Balakrishnan \etal \cite{jour/tmi/balakrishnan2019voxelmorph} proposed a learning-based framework for image registration.
The framework could extend to multi-modality images when the anatomical label is provided during the training. 
Furthermore, Luo \etal \cite{conf/miccai/luo2020mvmm} proposed a group-wise registration network, which could jointly register multiple atlases to the target image. 
Nevertheless, these methods required extensive anatomical labels for network training, which prevents them from unlabeled datasets.

At present, based on image-to-image translation generative adversarial network (GAN) \cite{conf/eccv/huang2018multimodal}, several unsupervised registration networks had been proposed.
Qin \etal \cite{conf/miccai/qin2019unsupervised} disentangled a shape representation from multi-modality images via GAN, then a convenient mono-modality similarity metric could be applied on the shape representation for registration network training. 
Arar \etal \cite{conf/cvpr/arar2020unsupervised} connected registration network with a style translator. 
The network could jointly perform spatial and style transformation on a moving image, and was trained by minimizing the difference between the transformed moving image and fixed image. 
The basic idea of these GAN-based methods is converting the multi-modality registration problem into a mono-modality one. Unfortunately, GAN methods easily cause geometrically distortions and intensity artifacts during image translation \cite{conf/cvpr/zhang2018translating}, which may lead to unrealistic registration results. 

In this work, we propose an end-to-end multi-modality 3D registration network (MMRegNet). 
The main contributions are:
(1) We present a spatially encoded gradient information (SEGI), which can provide a similarity criteria to train the registration network in an unsupervised manner.
(2) We evaluated our method on multi-modality cardiac left ventricle and liver registration tasks and obtained promising performance on both applications.

\section{Method}
\paragraph{\textbf{Registration Network:}}
Let $I_m$ and $I_f$ be a moving and fix image, respectively. Here, $I_m$ and $I_f$ are acquired via different imaging protocols, and are defined in a 3-D spatial domain $\Omega$. We construct the MMRegNet based on a U-shape convolution neural network \cite{conf/miccai/ding2020cross}, which takes a pair of $I_m$ and $I_f$ images as input, and predicts forward $U$ and backward $V$ dense displacement fields (DDFs) between them simultaneously. Therefore, MMRegNet is formulated as follows,
\begin{equation}
(U,V)=f_\theta(I_m,I_f)
\end{equation}	
where ${\theta}$ is the parameter of MMRegNet. Each voxel $x \in \Omega$ in $I_m$ and $I_f$ can be transformed by $U$ and $V$ as follows,
\begin{equation}
({I}_m\circ U)(x)=I_m (x+U(x)),
\end{equation}	
\begin{equation}
({I}_f \circ V)(x)=I_f (x+V(x)),
\end{equation}
where $\circ$ is spatial transformation operation, and ${I}_m \circ U$, ${I}_f \circ V$ denote the moved image of $I_m$, $I_f$, respectively.

	

\paragraph{\textbf{Spatially Encoded Gradient Information:}}
Generally, the parameter of  MMRegNet could be optimized by minimizing the intensity-based criteria, such as mean square error of intensity between the moved image and fixed image. However, such metrics are ill-posed when applied in multi-modality scenarios. This is because the intensity distribution of an anatomy usually varies across different modalities of images. Normalized gradient information (NGI) \cite{conf/miccai/haber2006intensity} is widely explored for conventional multi-modality registration methods. 
\begin{figure}[htb] 
	\centering
	\includegraphics[width=1.0\textwidth]{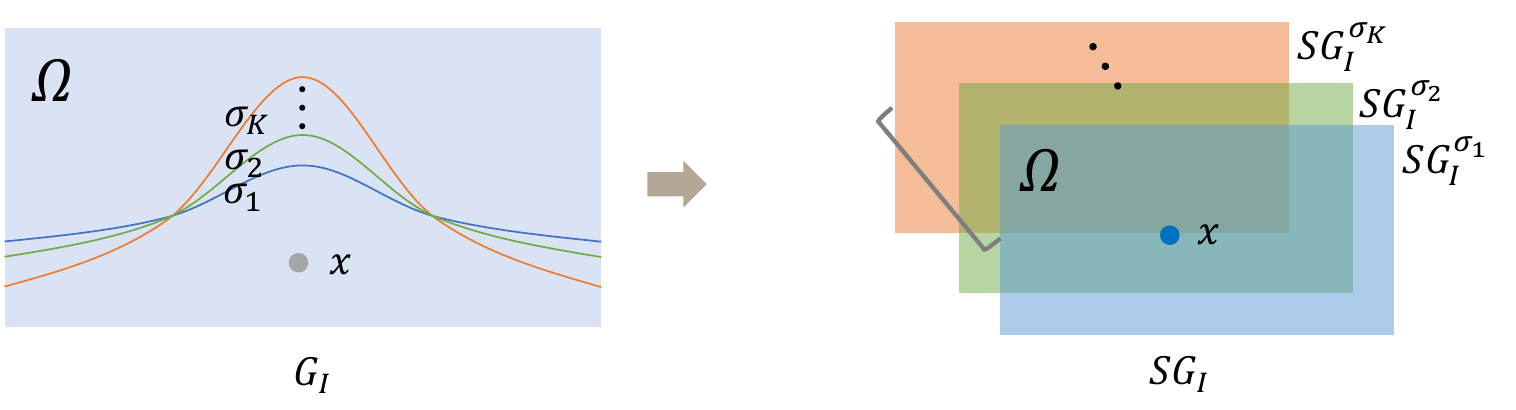}
	\caption{A visual demonstration of the SEGI. 
	}
	\label{fig:2}    
\end{figure}
The basic idea of NGI is based on the assumption that image structures can be defined by intensity changes. Let $G_{I}$ be the NGI of an intensity image $I$, each element of $G_{I}$ is calculated as follows,
\begin{equation}
G_{I}(x)={\frac{\nabla I(x)}{\|\nabla I(x)\|_2}},
\end{equation}
where $x\in \Omega$, and $\nabla I$ refers to the gradient of image $I$. Ideally, MMRegNet can be trained by minimizing the difference between $G_{I_m \circ U}$ and $G_{I_f}$. However, such a criteria is sensitive to the noises or artifacts of intensity images. It is still error-prone to train a registration network via NGI criteria in our practical experiments. To overcome this, we extend NGI to SEGI. Figure \ref{fig:2} illustrates the idea of SEGI, it is achieved by introducing a set of spatial variables ${\Sigma}=\{\sigma_1,\sigma_2\cdots,\sigma_K\}$ to the standard NGI ($G_I$). For each spatial variable $\sigma_k$, we compute its associated SEGI ($SG_I^{\sigma_k}$) as follows,
\begin{equation}
SG_{I}^{\sigma_k}(x)=\sum_{p\in \Omega}{\mathcal{N}(p|x,\sigma_k^2)  \frac{\nabla I(p)}{\|\nabla I(p)\|_2}},
\end{equation}
where $x \in \Omega$, and $\mathcal{N}(p|x,\sigma_k^2)$ denotes Gaussian distribution. Notably, we accumulate gradient information around $x$ for a more robust representation of the intensity change. 
Finally, given a set of spatial variables $\Sigma$, the SEGI of an intensity image $I$ is defined as,
\begin{equation}
SG_I=\{SG_{I}^{\sigma_1},SG_{I}^{\sigma_2},\cdots,SG_{I}^{\sigma_K}\}.
\label{eq:sg_variable}
\end{equation}

\paragraph{\textbf{Loss Function:}}
We train the network by minimizing the cosine distance between the SEGI of moved ($SG_{I_m \circ U}$) and fixed ($SG_{I_f}$) images,
\begin{equation}
\mathcal{L}_{SG}=\frac{1}{K}\sum_{k=1}^{K}\mathcal{D}({SG}_{I_m \circ U}^{\sigma_k},SG_{I_f}^{\sigma_k}),
\end{equation}
\begin{equation}
	\mathcal{D}({SG}_{I_m \circ  U }^{\sigma_k},SG_{I_f}^{\sigma_k})=\frac{-1}{|\Omega|}\sum_{x \in \Omega}cos({SG}_{I_m \circ U}^{\sigma_k}(x),SG_{I_f}^{\sigma_k}(x)),
	\label{eq:sg}
\end{equation}
where $|\Omega|$ counts the number of voxels in an image, and $cos(\bm{A},\bm{B})$ calculates the cosine distance between vector $\bm{A}$ and $\bm{B}$. 

Meanwhile, MMRegNet is designed to simultaneously predict $U$ and $V$ for each pair of $I_m$ and $I_f$. Normally, $U$ and $V$  should be inverse of each other. Hence, we employ the cycle consistent constraint \cite{conf/miccai/ding2020cross} for the DDFs such that each $I_m$ can be restored  to its original one after transforming by $U$ and $V$ in succession,
\begin{equation}
\mathcal{L}_{CC}=\frac{1}{|\Omega|}\sum_{x \in \Omega}\|{I}_{m} \circ U \circ V (x)-I_m(x)\|_1.
\end{equation}

Finally, the total trainable loss of the registration network is defined as follows,
\begin{equation}
\label{equ:loss}
\mathcal{L}=\mathcal{L}_{SG}+\lambda_1 \mathcal{L}_{CC}+{\lambda_2}\{\Psi(U)+\Psi(V)\},
\end{equation}
where $\Psi(U)$ and $\Psi(V)$ are smoothness regularization terms for DDFs, and $\lambda_1$, $\lambda_2$ are the hyper-parameters.

\section{Experiments and Results}
\paragraph{\textbf{Experimental Setups:}}
MMRegNet was implemented by the TensorFlow on an NVIDIA P100. We tested it on two public datasets, \ie, the MM-WHS\footnote{www.sdspeople.fudan. edu.cn/zhuangxiahai/0/mmwhs/} \cite{jour/mia/zhuang2019evaluation} and CHAOS\footnote{https://chaos.grand-challenge.org/} \cite{jour/mia/kavur2021chaos}.  
\begin{itemize}
\item MM-WHS: MM-WHS contains multi-modality (CT, MR) cardiac medical images. We utilized 20 MR and 20 CT images for left ventricle registration task. MMRegNet was trained to perform the registration of MR to CT images.  
\item CHAOS: CHAOS contains multi-modality abdominal images from healthy volunteers. For each volunteer, the dataset includes their T1, T2 and CT images. We adopted 20 T1 MR, 20 T2 MR and 20 CT images for liver registration. 
\end{itemize}
During the training phase, we employed ADAM optimizer to optimize the network parameters for 5000 epochs. 
The spatial variables $\Sigma$ were given to $\{1,1.5,3\}$ practically, aiming to capture multi-scale of robust gradient information for registration.
Meanwhile, we tested $\lambda_1 $ and $\lambda_2$ with four different weighting values, \ie, 0.01, 0.1, 1, 10. According to the corresponding results of different setups, we set $\{\lambda_1 = 0.1, \lambda_2 = 10 \}$ and $\{\lambda_1 = 0.1, \lambda_2 = 1 \}$ for MM-WHS and CHAOS dataset, respectively. 
 To evaluate the performance of MMRegNet, we computed the Dice (DS) and average symmetric surface distance (ASD) between the corresponding label of moved and fix images. All experimental results were reported by 4-fold cross-validation.
 
\paragraph{\textbf{Results:}}
We compared our registration method with three state-of-the-art multi-modality registration methods.
\begin{itemize}
	\item Sy-NCC: The conventional affine + deformable registration, which is based on the symmetric image normalization method with normalized cross-correlation (NCC) as optimization metric \cite{journal/inj/avants2009advanced}. We implemented it based on the popular ANTs software package\footnote{https://github.com/ANTsX/ANTsPy} . 
	\item Sy-MI: The Sy-NCC method which uses the MI instead of the NCC as optimization metric. 
	\item VM-NCC: The state-of-the-art registration network \cite{jour/tmi/balakrishnan2019voxelmorph}, which was trained by using the NCC as training criteria. We adopted their official online  implementation\footnote{https://github.com/voxelmorph/voxelmorph}.
	
\end{itemize}

\begin{table*}[tb]
	\centering
	\caption{The performance of different multi-modality registration methods on MM-WHS dataset.}
	
	\begin{tabular}{lllll}
		
		\hline
		\multicolumn{1}{l|}{\multirow{2}{*}{Method}} & \multicolumn{2}{c}{LVC (MR$\rightarrow$CT)}  & \multicolumn{2}{c}{Myo (MR$\rightarrow$CT)}   \\ 
		\cline{2-5} 
		\multicolumn{1}{l|}{}  & \multicolumn{1}{c}{DS (\%)$\uparrow$} & \multicolumn{1}{c|}{ASD (mm)$\downarrow$} & \multicolumn{1}{c}{DS (\%)$\uparrow$} & 
		\multicolumn{1}{c}{ASD (mm)$\downarrow$} \\ 
		\cline{1-5}

		\multicolumn{1}{l|}{Sy-NCC \cite{journal/inj/avants2009advanced}}  &70.07$\pm$16.57& \multicolumn{1}{l|}{4.51$\pm$2.67}  & 50.66$\pm$16.02 &    \multicolumn{1}{l}{4.10$\pm$1.77}  \\
		
		\multicolumn{1}{l|}{Sy-MI \cite{journal/inj/avants2009advanced}}  &69.16$\pm$15.25 & \multicolumn{1}{l|}{4.66$\pm$2.54}  & 49.00$\pm$16.21 &    \multicolumn{1}{l}{4.34$\pm$2.04}  \\
		
		\multicolumn{1}{l|}{VM-NCC \cite{jour/tmi/balakrishnan2019voxelmorph}}  &79.46$\pm$8.73 & \multicolumn{1}{l|}{\textbf{2.81$\pm$1.05}}  & 62.77$\pm$9.51 &    \multicolumn{1}{l}{\textbf{2.49$\pm$0.61}}  \\
		
		
		
		\multicolumn{1}{l|}{MMRegNet}    & \textbf{80.28$\pm$7.22} & \multicolumn{1}{l|}{{3.46$\pm$1.30}}  & \textbf{62.92$\pm$8.62}&    \multicolumn{1}{l}{{3.01$\pm$0.74}} \\
		\hline
		
	\end{tabular}
	\label{tab:mmwhs}
\end{table*}

\begin{table*}[tb]
	\centering
	\caption{The performance of different multi-modality registration methods on CHAOS dataset.}
	
	\begin{tabular}{lllll}
		
		\hline
		\multicolumn{1}{l|}{\multirow{2}{*}{Method}} & \multicolumn{2}{c}{Liver (T1$\rightarrow$CT)}  & \multicolumn{2}{c}{Liver (T2$\rightarrow$CT)}   \\ 
		\cline{2-5} 
			\multicolumn{1}{l|}{}  & \multicolumn{1}{c}{DS (\%)$\uparrow$} & \multicolumn{1}{c|}{ASD (mm)$\downarrow$} & \multicolumn{1}{c}{DS (\%)$\uparrow$} & 
	\multicolumn{1}{c}{ASD (mm)$\downarrow$} \\ 
		\cline{1-5}

		\multicolumn{1}{l|}{Sy-NCC \cite{journal/inj/avants2009advanced}}  &74.94$\pm$11.05 & \multicolumn{1}{l|}{8.46$\pm$4.10}  & 75.46$\pm$9.42 &    \multicolumn{1}{l}{8.41$\pm$3.86}  \\
		
		\multicolumn{1}{l|}{Sy-MI \cite{journal/inj/avants2009advanced}}  &73.88$\pm$10.08 & \multicolumn{1}{l|}{8.84$\pm$3.70}  & 75.82$\pm$7.23 &    \multicolumn{1}{l}{8.32$\pm$2.73}  \\
		
		\multicolumn{1}{l|}{VM-NCC \cite{jour/tmi/balakrishnan2019voxelmorph}}  &74.63$\pm$6.54 & \multicolumn{1}{l|}{8.25$\pm$2.17} & 71.10$\pm$6.09 &    \multicolumn{1}{l}{9.30$\pm$2.01}  \\
		
		
		
		\multicolumn{1}{l|}{MMRegNet}    & \textbf{79.00$\pm$8.06} & \multicolumn{1}{l|}{\textbf{7.03$\pm$2.55}}  & \textbf{76.71$\pm$8.80}&    \multicolumn{1}{l}{\textbf{7.87$\pm$1.75}} \\
		\hline
		
	\end{tabular}
	\label{tab:chaos}
\end{table*}

\begin{figure}[htb] 
	\centering
	\includegraphics[width=1.0\textwidth]{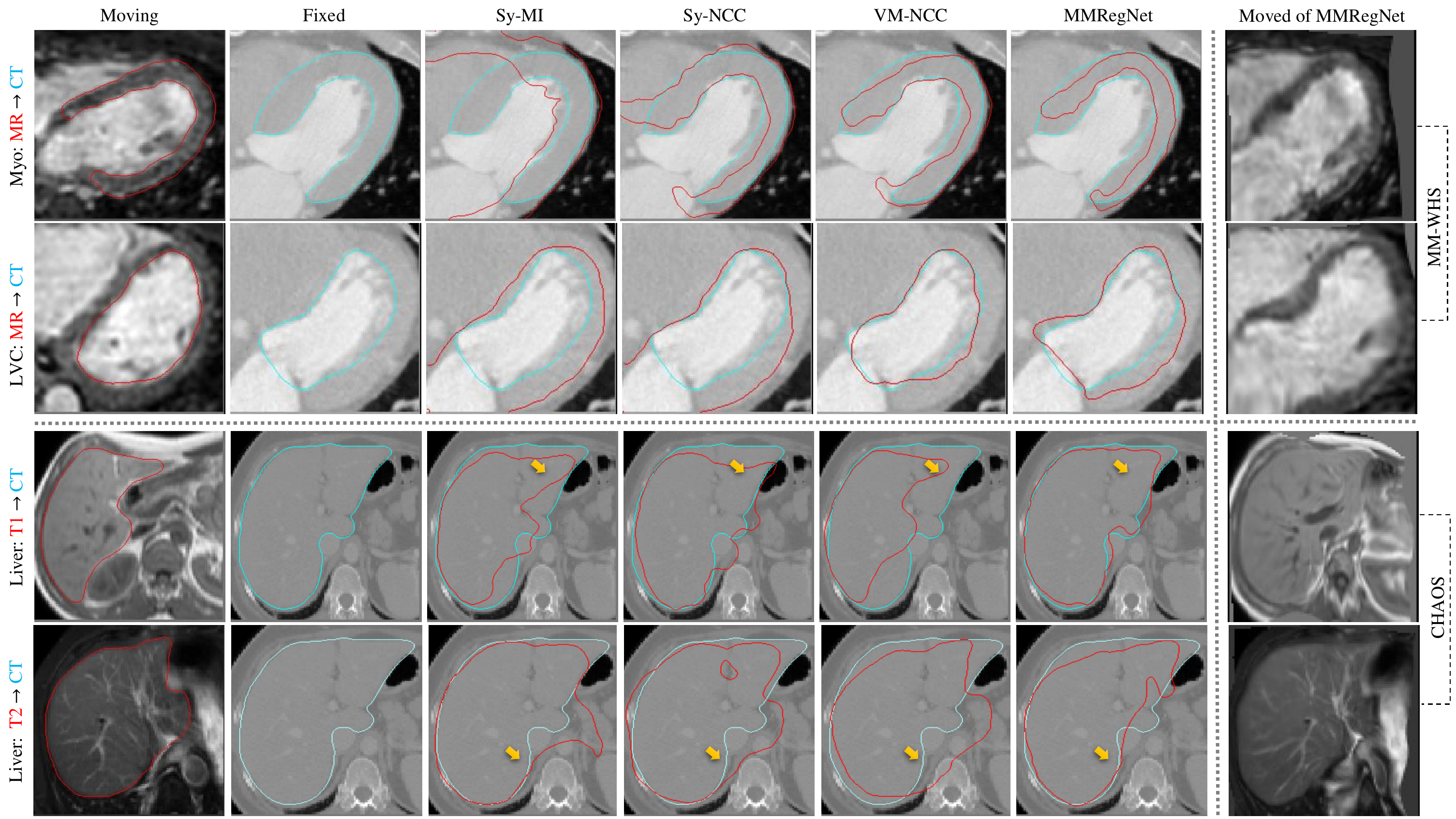}
	\caption{Visualization of different methods on MM-WHS and CHAOS datasets. The showed images are the representative cases in terms of DS by MMRegNet. The blue contours are the gold standard label of the fixed images, while the red contours delineate the label of moving or moved images. We indicate the advantage of MMRegNet via yellow arrows. Moreover, the last column presents the moved images of MMRegNet. (The reader is referred to the online version of this article)}
	\label{fig:mmwhse_chaos}    
\end{figure}

Table \ref{tab:mmwhs} shows the results on MM-WHS dataset. 
Compared with the conventional methods (Sy-NCC and Sy-MI), MMRegNet could achieve better performance on both left ventricle cavity (LVC) and left ventricle myocardium (Myo). Notably, compared to the state-of-the-art registration network, i.e., VM-NCC, MMRegNet obtained comparable results in terms of DS and ASD. 
This reveals that MMRegNet is applicable for multi-modality registration tasks, and the proposed SEGI could serve as another efficient metric, such as MI and NCC, for multi-modality registration.

Table \ref{tab:chaos} shows the results on CHAOS dataset. 
We independently reported the registration result of T1 or T2 to CT images.  MMRegNet achieved comparable accuracy to the state-of-the-art conventional methods, \ie, Sy-MI and Sy-NCC. Meanwhile, compared to VM-NCC, MMRegNet obtained average 4.99\%  (T1$\rightarrow$CT: 4.37\%, T2$\rightarrow$CT: 5.61\%) and 1.33 mm (T1$\rightarrow$CT: 1.22 mm, T2$\rightarrow$CT: 1.43 mm) improvements in terms of DS and ASD, respectively.  This indicates that MMRegNet could achieve promising performance for multi-modality registration tasks.      

Additionally, Figure \ref{fig:mmwhse_chaos} visualizes four representative cases from the two datasets. On MM-WHS dataset, one can observe that both  VM-NCC and MMRegNet achieved better visual results than  Sy-MI and Sy-NCC, which is consistent with the quantitative results in Table \ref{tab:mmwhs}. On CHAOS dataset, the yellow arrows highlight where MMRegNet could obtain relative reasonable results than other methods.  


\section{Conclusion}
In this paper, we present an end-to-end network for multi-modality registration. 
The network is both applicable for heart and liver registration tasks. 
Meanwhile, we propose SEGI to obtain a robust structural representation for multi-modality images, and then applied it as the loss function for unsupervised registration network training. 
The results showed that MMRegNet could achieve promising performance when comparing with the state-of-the-art registration methods. Further work will extend MMRegNet to other multi-modality datasets.
	
\bibliographystyle{splncs04}
	
\bibliography{ref}

\end{document}